\documentclass[ reprint, amsmath,amssymb,apl,aip]{revtex4-2}
\usepackage{color}
\usepackage{verbatim}
\usepackage{lipsum}
\usepackage{hyphenat}
\usepackage{dirtytalk}
\usepackage{graphicx}% Include figure files
\usepackage{dcolumn}% Align table columns on decimal point
\usepackage{bm}% bold math
\usepackage{tabularx}
\usepackage{hyperref}% add hypertext capabilities
\usepackage{mathtools}
\usepackage{subfigure}
\usepackage{tabularx}
\usepackage{etoolbox}
\usepackage{colortbl}
\usepackage{multirow}
\preprint{APS/123-QED}
\newsavebox{\measurebox}

\newcolumntype{l}{X}
\newcolumntype{s}{>{\hsize=0.3\hsize}X}
\newcolumntype{m}{>{\hsize=0.85\hsize}X}

\begin{document}

\title{Precision Multi\hyp{}Mode Microwave Spectroscopy of Paramagnetic and Rare-Earth Ion Spin Defects in Single Crystal Calcium Tungstate}
\author{Elrina Hartman}
 \affiliation{Quantum Technologies and Dark Matter Labs, Department of Physics, University of Western Australia, 35 Stirling Highway, Crawley, WA 6009, Australia.}
\email{To whom correspondence should be addressed: michael.tobar@uwa.edu.au \ \ elrina.hartman@research.uwa.edu.au}
\author{Michael E Tobar}
\affiliation{Quantum Technologies and Dark Matter Labs, Department of Physics, University of Western Australia, 35 Stirling Highway, Crawley, WA 6009, Australia.}
\author{Ben T McAllister}%
\affiliation{Quantum Technologies and Dark Matter Labs, Department of Physics, University of Western Australia, 35 Stirling Highway, Crawley, WA 6009, Australia.}
\author{Jeremy Bourhill}
\affiliation{Quantum Technologies and Dark Matter Labs, Department of Physics, University of Western Australia, 35 Stirling Highway, Crawley, WA 6009, Australia.}
\author{Maxim Goryachev}
\affiliation{Quantum Technologies and Dark Matter Labs, Department of Physics, University of Western Australia, 35 Stirling Highway, Crawley, WA 6009, Australia.}

\date{\today}

\begin{abstract}

We present experimental observations of dilute ion spin ensemble defects in a low\hyp{}loss single crystal cylindrical sample of CaWO$_4$ cooled to $30$ mK in temperature. Crystal field perturbations were elucidated by constructing a dielectrically loaded microwave cavity resonator from the crystal. The resonator exhibited numerous whispering gallery modes with high $Q$\hyp{}factors of up to $3\times 10^7$, equivalent to a loss tangent of $\sim 3\times 10^{-8}$. The low loss allowed precision multi\hyp{}mode spectroscopy of numerous high $Q$\hyp{}factor photon\hyp{}spin interactions. Measurements between 7 to 22 GHz revealed the presence of Gd$^{3+}$, Fe$^{3+}$, and another trace species, inferred to be rare\hyp{}earth, at concentrations on the order of parts per billion. These findings motivate further exploration of prospective uses of this low\hyp{}loss dielectric material for applications regarding precision and quantum metrology, as well as tests for beyond standard model physics. 

\end{abstract}
\maketitle

Scheelite, or calcium tungstate (CaWO$_{4}$), has seen a significant rise in interest for its potential of application in a myriad of contexts from test of fundamental physics, communications, and quantum computing  when considered in conjunction with spin ensemble dopants such as Gd$^{3+}$, Er$^{3+}$\cite{PhysRev.168.370,Bernal2003,Bertaina2007,PhysRevLett.103.226402,mr-1-315-2020,PhysRevB.106.144412,sciadv.abj9786,Ourari2023}, Yb$^{3+}$ \cite{PhysRevB.79.172408}, Nd$^{3+}$ \cite{Garrett2004} and Pr$^{3+}$\cite{Aizenberg66,CAVALLI2016450,PhysRevA.107.022802}, with particular interest in utilizing rare\hyp{}earth ions for quantum information storage and processing, and on\hyp{}chip sensing\cite{PhysRevApplied.19.024067,PhysRevApplied.18.014054}. The scintillating dielectric also plays a key role in detection of rare events such as neutrinoless double $\beta$\hyp{}decay \cite{ZDESENKO2005249}, radioactive decay of very long\hyp{}living isotopes \cite{PhysRevC.70.064606} and searches for weakly interacting massive particles (WIMPs), a candidate for Dark Matter (DM) \cite{ANGLOHER2005325,SIVERS20121843}. The question of interest is whether or not either or both of these application avenues can be realised in a single crystal architecture at low temperatures, and what effects do the  residual impurity spin ensemble defects display. To that end, this report presents the findings of precision multi\hyp{}mode microwave electron spin resonance (ESR) spectroscopy, as developed in \cite{farr2013ultrasensitive}, to identify the paramagnetic residual impurities present in a high\hyp{}purity CaWO$_4$ sample that was purchased from SurfaceNet. 

\begin{figure}[t!]
\includegraphics[height=7.4cm]{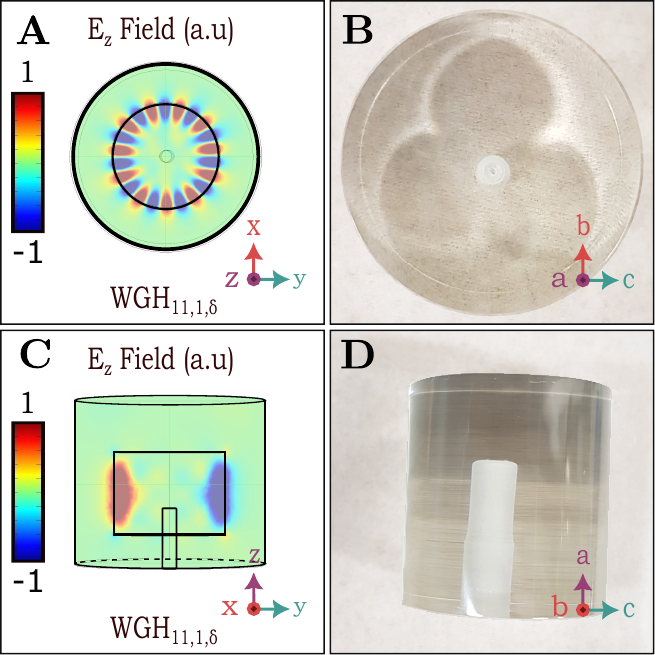}
\caption{The CaWO$_4$ crystal was grown along the $\langle100\rangle$ axis so that the c\hyp axis aligned with the cylinder radial\hyp axis, with a central hole drilled on one face of the cylinder for the purpose of mounting the crystal on a post. Here (\textbf{A}) and  (\textbf{C}) show the schematics of the dielectrically loaded resonator, and the electric\hyp{}energy density of the E$_z$ field of the quasi\hyp{}TM WGH$_{11,1,\delta}$ mode, calculated using COMSOL. Corresponding top (\textbf{B}) and side view (\textbf{D}) photographs of the dielectric sample are also shown. The borehole that appears white in the photograph is in a region of low field density for a WGM so frequency perturbations due to the presence of the sapphire post are negligible.}
\label{fig:sample}
\end{figure}

The manufacturer grew the sample in the $\langle100\rangle$ orientation such that the a\hyp{}axis of the crystal unit cell was aligned with the z\hyp{}axis of the macroscopic cylindrical geometry. The sample was then placed into a metallic cavity to form a dielectrically loaded cavity resonator as shown in Fig. \ref{fig:sample}. ESR studies allowed us to measure the crystal field parameters, Land\'e  g factors ($g_L$), zero field splittings (ZFS), and coupling rate (g), allowing the defect ion concentration to be calculated.
\begin{figure}[t!]
\includegraphics[height=7.5cm]{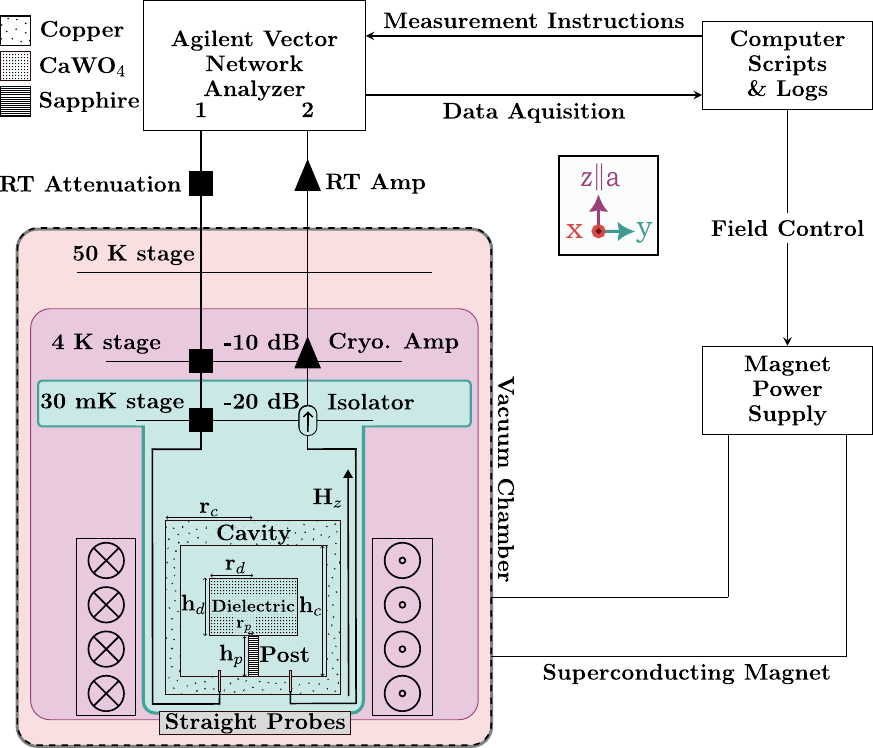}
\caption{The CaWO$_4$ crystal ($r_{d}=14.68\,$mm, $h_{d}=20.00\,$mm) was mounted on a sapphire post ($r_{p}=1.75\,$mm, $h_{p}=14.5\,$mm) clamped inside an oxygen\hyp free copper cavity ($r_{c}=25.0\,$mm, $h_{c}=40.0\,$mm), cooled to $30$ mk using a dilution refrigerator. A Vector Network Analyser (VNA) referenced by a H\hyp{}maser probed the microwave resonances, automated via computer scripts that set the parameters such as scan rate, power input, magnetic field strength, and frequency span.}
\label{fig:Setup}
\end{figure}
\begin{figure}[t!]
    \centering
\includegraphics[width=\linewidth]{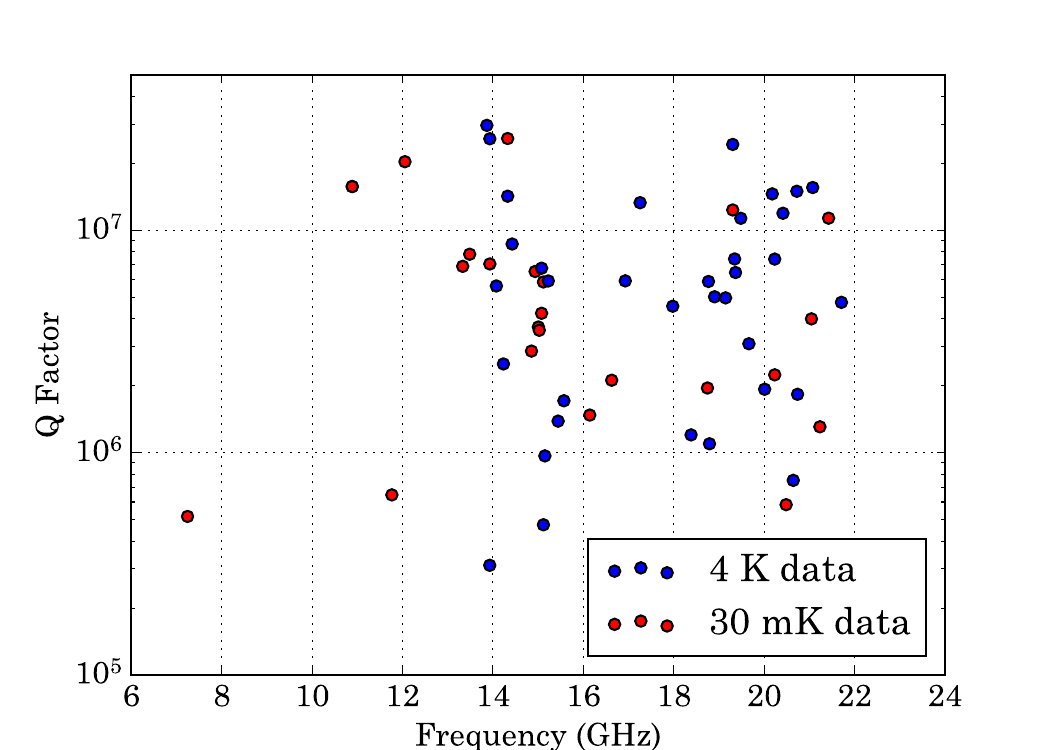}
    \caption{This sample exhibits high mode density. Shown is the zero\hyp{}field distribution of high\hyp{}$Q$ resonances at liquid helium temperature (blue). A spread of resonances were selected and remeasured at 30 mK (red), with additional lower\hyp{}$Q$ modes at lower frequencies for resolving ZFS.}
    \label{fig:qfactors}
\end{figure}
The dielectrically loaded cavity was suspended in a superconducting magnet bore at the mixing chamber plate, reaching a base temperature of $30$ mK when loaded in a dilution refrigerator as shown in Fig. \ref{fig:Setup}. Two coaxial antennae were inserted at the base of the cavity to allow excitation of high\hyp{}$Q$ Whispering Gallery Modes (WGM)\cite{wgmintro}. Further coaxial cabling connected the antennae to a room\hyp{}temperature Vector Network Analyser (VNA) with a high stability H\hyp{}maser as the frequency reference, for microwave excitation and readout. The probes were inserted to a depth adjusted for under\hyp{}coupling so that the intrinsic losses of the dielectric could be measured\cite{lowcoupling}. Input power was attenuated at the $4$ K and $30$ mK stage to reduce thermal noise from the input port. The fridge was first cooled to 4 K to assess the mode population in the crystal. Specific modes were not identified within the spectrum, however, note that WGMs are all characteristically high\hyp{}$Q$ ($Q>10^5$) \cite{farr2013ultrasensitive,Hatzon24,McAllister24}, which distinguishes them from both spurious and copper cavity modes. Furthermore, when conducting finite element modelling of the system in COMSOL Multiphysics, we find that the only high\hyp{}$Q$ modes predicted are WGMs. This is due to their coherent field distribution, an example of which can be seen in Fig. \ref{fig:sample}. A subset of these modes was then remeasured at 30 mK with trade-off considerations made between highest $Q$\hyp{}factor and evenly spaced frequency intervals as selection criteria. Between $4$ K and $30$ mK there was a small frequency shift and slight reduction in $Q$\hyp{}factor. Both effects had negligible impact on the sensitivity of the WGM ESR. A far more consequential temperature effect was the increase in spin population occupying the ground state at $30$ mK. A DC magnetic field was applied along the z\hyp{}axis of the cylindrical sample and ramped slowly in steps of $1$ mT, tuning the spin transition via the Zeeman effect, so they would interact with high\hyp{}$Q$ WGMs. Transmission spectra of each mode were recorded at all magnetic field values so the interactions could be captured, including impacts on the photon mode's frequency, $Q$\hyp{}factor, and transmission power. The forward power gain transfer function, specified by the scattering\hyp{}parameters ($S_{21}$), was measured via the VNA and modelled as a Breit\hyp Wigner distribution or asymmetric Fano resonance, given by \cite{Limonov2017};
\begin{equation}
   |S_{21}| =  1-\frac{(q\Gamma_{p}/2+\Delta)^2}{(\Gamma_{p}/2)^2+\Delta^2}\,,
\end{equation}
where \textit{q} is the Fano parameter measuring the ratio of resonant scattering to background scattering, $\Gamma_{p}$ is the photonic resonance line\hyp width, and $\Delta$ is the frequency detuning ($f-f_0$). The $Q$\hyp{}factors of the modes can be calculated from the line\hyp widths and are shown for zero DC magnetic field in Fig.  \ref{fig:qfactors}. The best $Q$\hyp{}factors in the crystal are of order of $3\times10^7$ close to 14 GHz, indicating a loss tangent of order tan$\delta\approx3\times10^{-8}$. The exact values of the anisotropic loss components could be verified more precisely using the WGM technique \cite{tobar98,krupka1999b,krupka99,farr2013ultrasensitive} in a different sample, which has the cylindrical z\hyp{}axis aligned with the c\hyp{}axis in future studies.

\begin{figure}[ht]
\centerline{\begin{minipage}{\linewidth}
\centering
\includegraphics[width=\textwidth]{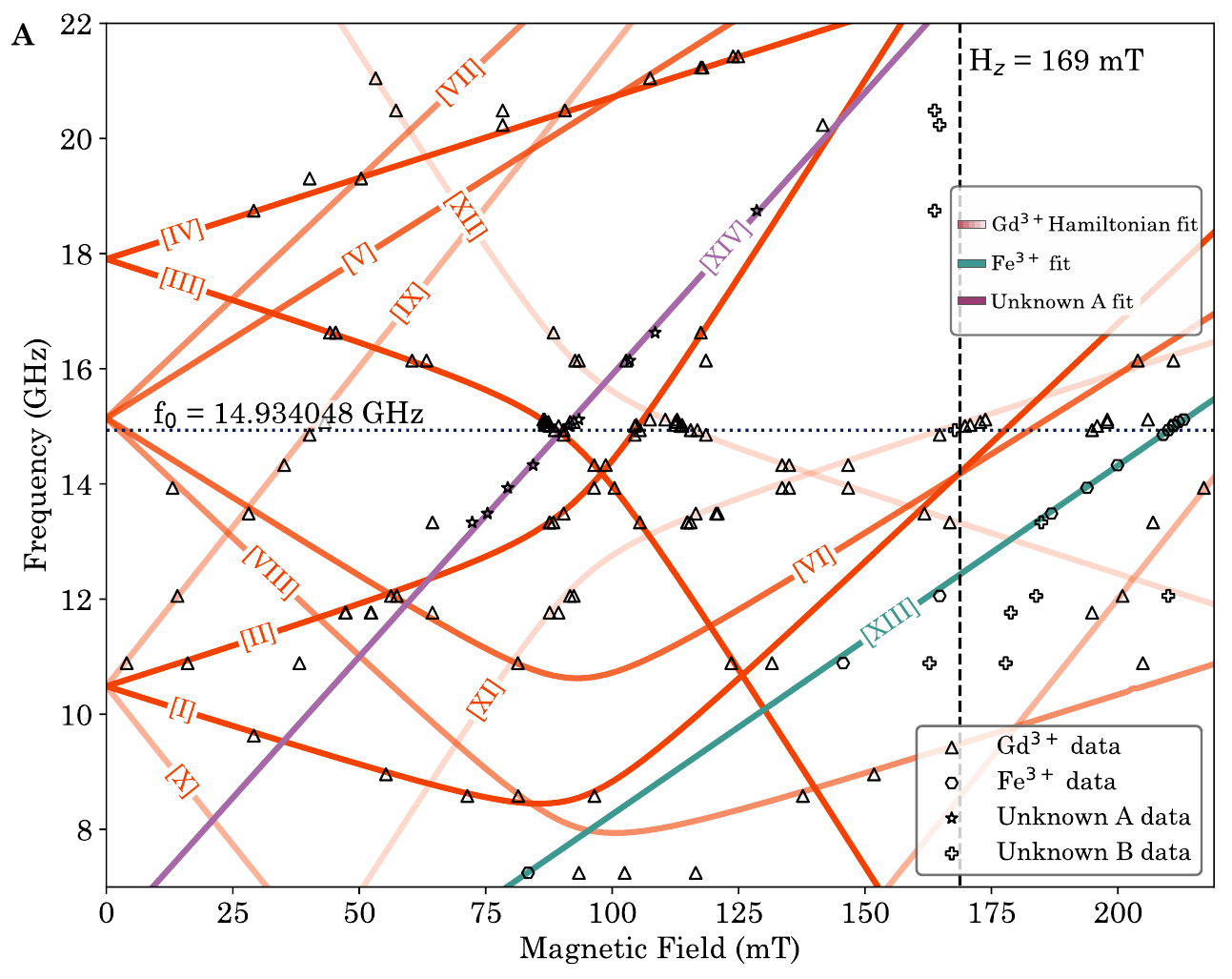}
\includegraphics[width=\textwidth]{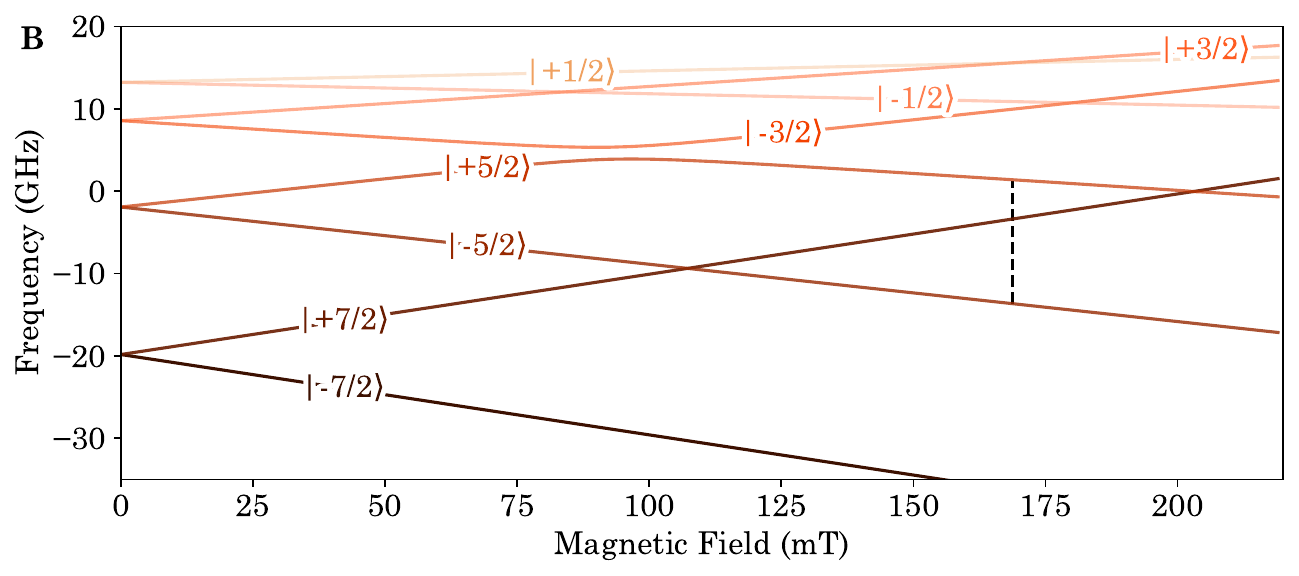}
\caption{\textbf{A} Shown are the perturbation sites in the frequency to magnetic field parameter space across the span of all WGMs employed. Spin transitions in CaWO$_4$:Gd$^{3+}$ (orange curves, labelled I to XII) are derived from the spin Hamiltonian (see equation \ref{eq:gdhamiltonian}). The Fe$^{3+}$ transition is shown in green ($g_L = 4.3$, ZFS $=2.20$ GHz, labelled XIII) and another unidentified paramagnetic impurity defect species ($g_L = 7$, ZFS $=6.1$ GHz, labelled XIV) is shown in purple and denoted ``Unknown A" data. Additional perturbations with no clear regression\hyp{}line  can be seen at higher field and are designated as ``Unknown B" data. \textbf{B} Plotted is the CaWO$_4$:Gd$^{3+}$ energy\hyp{}level diagram in units of frequency (GHz). The high\hyp{}$Q$ WGM used to calculate coupling strength and spin concentration is highlighted with the horizontal dotted line at $14.934048$ GHz and the vertical dashed line illustrates correspondence of the magnetic field location ($169$ mT) in \textbf{A} and the spin transition in \textbf{B}. The frequency perturbation of this spin\hyp{}photon interaction is shown in more detail in Fig. \ref{closeup}.}
\label{fig:SpecResults}
\end{minipage}}
\end{figure}

In order to measure the spin transitions within the crystal, we excited them using multiple high\hyp{}$Q$ WGMs between $7$ to $22$ GHz. When the energy of a spin transition matches the energy of an injected WGM, they interact allowing the spin transition to be observed \cite{Abe2011,PhysRevA.85.053806}. The modes were tracked simultaneously as the magnetic field strength was increased. The resulting mode\hyp{}map is plotted in Fig. \ref{fig:SpecResults}, and as will be discussed in more detail below, $g_L$ and ZFS were extracted from this data (see Tab. \ref{tbl:Detune}). These two parameters take unique values for a paramagnetic defect in a given host, and thus comparing observed $g_L$ and ZFS values to previous measurements of doped CaWO$_4$ allowed for the identification of some of the paramagnetic impurities within our crystal.

The energy of the spin ensembles are modelled using a spin Hamiltonian that takes into consideration the Zeeman splitting, the crystal field, and the hyperfine splitting due to spin\hyp{}orbit coupling. On the scale of GHz, spin\hyp{}orbit coupling was not elucidated by the data, as the detuning resolution and the uncertainty of the detuning from the fitting is on the same order of magnitude as the hyper\hyp{}fine structure constants. Thus, their contribution was neglected for the present study and our focus was to determine the crystal field zero\hyp{}field splitting and the Zeeman splitting term. The value of $g_L$ was extracted from the slopes of the Zeeman levels as a function of magnetic field. The crystal field parameters were constructed by well\hyp{}defined tesseral harmonics that describe the spin energy levels. We chose the Stevens representation, which implements a spin operator basis, and determines the Stevens coefficients experimentally. Comparison between Fig. \ref{fig:SpecResults} data and Hamiltonian fits from prior literature, served to check, corroborate or eliminate potential impurity species.

The scheelite unit cell comprises two types of atom clusters, O\hyp{}Ca and O\hyp{}W. Whilst the the global symmetry is tetragonal, the Ca and W centered clusters present tetragonal and distorted hexagonal symmetries respectively, by the arrangement of surrounding oxygen. The tetragonal Ca cluster has symmetry isomorphic to S$_{4}$ point symmetry \cite{NAZAROV2004291} and is a likely occupation site for replacement by rare\hyp{}earth and
shell\hyp{}3d transition metal impurities. Through the process of matching $g_L$, ZFS and other crystal field parameters we found that Fe$^{3+}$ and Gd$^{3+}$ were present in the sample, along with other unidentified trace rare\hyp{}earth contaminants.
\begin{table}[!h]
\centering
 \begin{tabularx}{0.45\textwidth}{msssl}%{@{} *4{>{\centering\arraybackslash}X}@{}}%{bbbbbbb}
\hline
\centering Species&$\Delta S_z$&Line&ZFS (GHz)&\centering\arraybackslash Transition\\
\hline
\hline
&&&&\\
CaWO$_{4}$:Gd$^{3+}$&$1$&I&\multirow{2}{*}{$10.49$}& $|+5/2\rangle\rightarrow |+3/2\rangle$\\
$g_L=1.99$&&II&&$|-5/2\rangle \rightarrow|-3/2\rangle$\\
&&III&$\multirow{2}{*}{17.90}$&$|+7/2\rangle\rightarrow|+5/2\rangle$\\
&&IV&&$|-7/2\rangle\rightarrow|-5/2\rangle$\\
&$2$&V&$\multirow{2}{*}{15.14}$& $|-5/2\rangle\rightarrow|-1/2\rangle$\\
&&VI&&$|+5/2\rangle\rightarrow|+1/2\rangle$\\
&$3$&VII&\multirow{2}{*}{$15.14$}&$|-5/2\rangle\rightarrow|+1/2\rangle$\\
&&VIII&&$|+5/2\rangle\rightarrow|-1/2\rangle$\\
&$4$&IX&\multirow{2}{*}{$10.49$}& $|-5/2\rangle\rightarrow|+3/2\rangle$\\
&&X&& $|+5/2\rangle\rightarrow|-3/2\rangle$\\
&$5$&XI& $0.0$ & $|-5/2\rangle\rightarrow|+5/2\rangle$\\
&&XII& $28.33$ &$|+7/2\rangle\rightarrow|-3/2\rangle$\\
\hline\\
CaWO$_4$:Fe$^{3+}$ \\ 
$g_L=4.3$& - &XIII& $2.20$ & -\\
\hline\\
\vspace{0.01cm}
Unknown A& -&XIV & $6.10$ & -\\
$g_L=7$&&&\\
\hline
\end{tabularx}
    \caption{Properties of spin transitions determined from the multi\hyp{}mode spectroscopy displayed in Fig. \ref{fig:SpecResults}A. Here $\Delta S_{z}$ is the change in spin quantum number.}
\label{tbl:Detune}
\end{table}
The crystal field of CaWO$_4$:Gd$^{3+}$ was modelled in this work by the following Hamiltonian;
\begin{equation}
    \begin{split}
        \mathcal{H} = \,& g_{L} \mu_{B} H_{z} S_{z} + B^0_2 O^0_2+ B^0_4 O^0_4\\
    & + B^4_4 O^4_4+ B^0_6 O^0_6+ B^4_6 O^4_6 \,,
    \end{split}
    \label{eq:gdhamiltonian}
\end{equation}
where $O^j_k$ are Stevens operators, $S=\frac{7}{2}$ and $g_{L}=1.99$. The crystal field parameters are; $B^0_2=-9.215\cdot 10^{-1}$, $B^0_4=-1.139\cdot 10^{-3}$, $B^4_4=-7.015\cdot 10^{-3}$, $B^0_6=5.935\cdot 10^{-7}$
, and $B^4_6=4.747\cdot10^{-7}$ %(also equal to lit.)
in units of GHz. This Hamiltonian allows the determination of the Zeeman energy levels of Gd$^{3+}$ in CaWO$_4$, which are shown in Fig. \ref{fig:SpecResults}\textbf{B}. Note, not all the energy levels are occupied or coupled to with the present setup, thus, not every energy level theoretically predicted is reflected in the experiment. The energy difference (vertical dashed line in Fig. \ref{fig:SpecResults}\textbf{B}) between the allowed transitions of these levels (coloured lines) allows one to identify the spin transitions in Fig. \ref{fig:SpecResults}\textbf{A}. This result is in good agreement with the literature \cite{Baibekov_2017} with the only discrepancy being a $3.14\%$ increase in the value of $B^0_2$ which dictates the ZFS values. This is due to the difference in orientation of this sample's c\hyp{}axis to that of Baibekov et.al. who use a $\langle001\rangle$ cut single crystal. Gd$^{3+}$ exhibits a distinctive electronic structure with the ground and first excited state are separated by a seven photon transition \cite{YVO}. Occupation of states requiring a high photon number to transition to are less likely to occur. Hence, spin occupation is distributed to a greater number of excited states than would be expected at thermodynamic equilibrium when cooled to $30$ mK. Time series data verify that the tuning rate of the magnetic field was sufficiently slow to avoid reading transient and heating effects at the instance of data acquisition. The results presented here are consistent with the findings in \cite{YVO} where this was indeed the case. To speculate at the source of this Gd$^{3+}$ contamination, we note that the YVO$_4$ sample reported in \cite{YVO} also had excess Gd$^{3+}$ ions, and is purchased from this same manufacturer, and it may be that one of their production stations was contaminated with gadolinium. The presence of Gd$^{3+}$ may enhance both photoluminescence and thermal\hyp{}luminescence of the crystal, benefitting potential bolometric and scintillating applications.

The Fe$^{3+}$ ($g_L=4.3$, ZFS $= 2.20$ GHz) transition was observed \cite{Fe3_1,MCGAVIN1985321}. Coupling to this species was too weak to fit to and hence determine the concentration. Fe$^{3+}$ spin ensembles have been shown to have long coherence lengths \cite{fe3+spinqubits} and are a promising candidate for pulsed qubit manipulations in scheelite. Another clear linear regression is present at low frequency and high magnetic field, denoted as ``Unknown A". This unknown spin ensemble has low ZFS and high $g_L$ factor, indicating that it may be a rare\hyp{}earth species. While systematic comparison of measured and literature crystal field parameters and g factors was sufficient to identify Gd$^{3+}$ and Fe$^{3+}$, more information will be required to confirm the identity of Unknown A. Consideration of natural abundances and further experimental investigation using other forms of spectroscopy will inform the literature review required for likely candidates. Most rare\hyp{}earth dopants in scheelite and scheelite\hyp{}like hosts have been characterised using a plethora of spectroscopic techniques such as Raman spectroscopy, X\hyp{}ray Powder Diffraction, Absorbance and Emission Spectroscopy, VUV synchrotron emission, mass spectrometry, and laser ablation. This field of research is resource rich due to continued and growing interest in rare\hyp{}earth enhancement of phosphor technologies such as La$^{3+}$, Ce$^{3+}$, Sm$^{3+}$, and Tb$^{3+}$ use in scintillators, LED's, and bio\hyp{}sensing or Er$^{3+}$ and Nd$^{3+}$ use in solid\hyp{}state lasers. Therefore, efforts to identify Unknown A are ongoing, but in a manner that prioritises preserving the single\hyp{}crystal sample so as to push the frontier of these rare\hyp{}earth applications in the dilute case. Other residual perturbation sites, simply designated as ``Unknown B" are shown with no clear trend or regression to allow identification. More data in this region of the perturbation map is need to elucidate these contributions to the spectra.

\begin{figure}[t!]
    \centering
	\includegraphics[width=\linewidth]{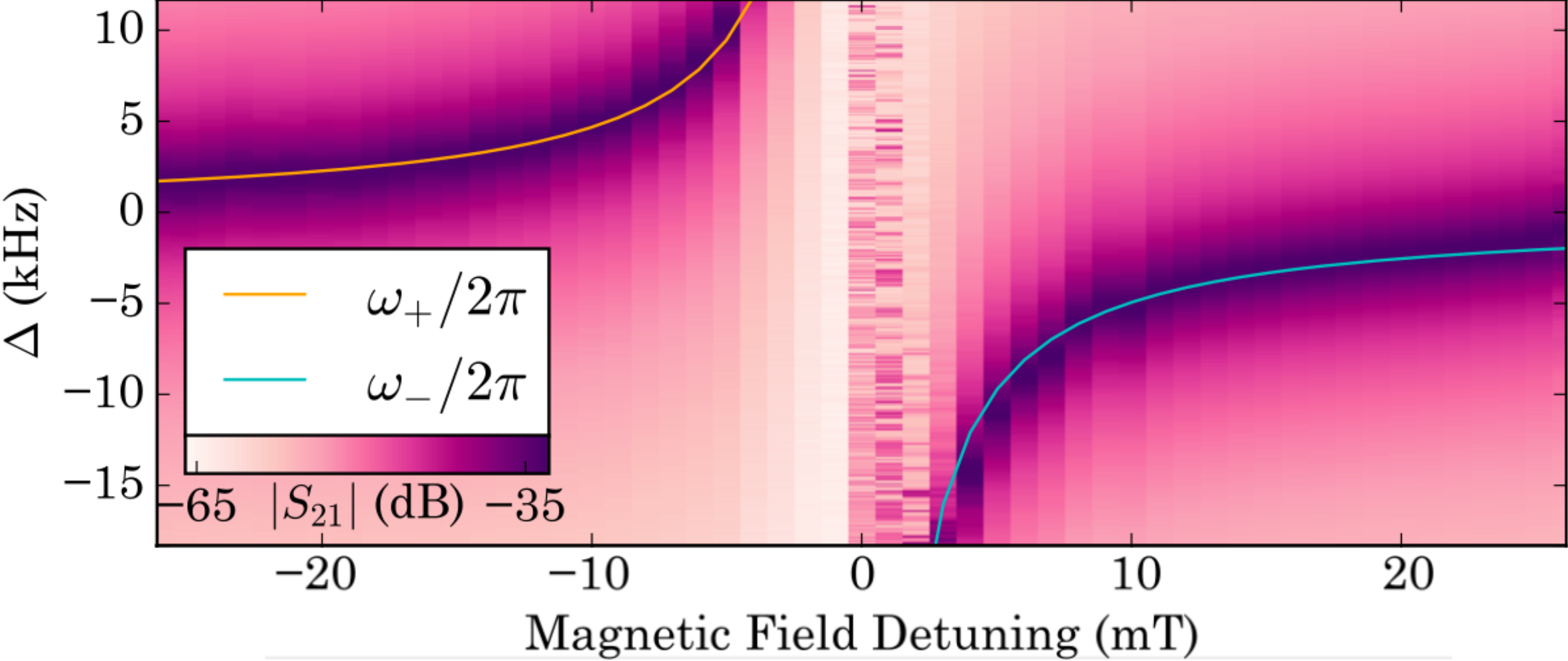}
	\caption{Density plot of the mode interaction between a Gd$^{3+}$ defect transition, and the $14.934048$ GHz WGM at 169 mT. The fit from Eq. (\ref{eq:freq detune}) is overlayed allowing calculation of the coupling between the photon and the spin of, $g=1.12\pm0.34$ MHz, with a measured $Q$-factor of the WGM of $6.5\times10^6$ and half bandwidth of $\Gamma_p/2=1.14$ kHz, so $\Gamma_p<<2g$.}
	\label{closeup}
\end{figure}

To estimate the concentration, we analysed an interaction between a Gd$^{3+}$ ion defect transition, and the 14.934048 GHz WGM at 169 mT, as highlighted in Fig. \ref{fig:SpecResults} and shown in more detail in Fig. \ref{closeup}, which may be modelled as a coupled harmonic oscillator, with $\omega_+$ and $\omega_{-}$ as the normal mode eigensolutions to the characteristic equation \cite{83834,PhysRevB.96.045141,tobar2023gravitational,Jeremy2016};
\begin{equation}
        \omega_{\pm}=\frac{1}{\sqrt{2}}\sqrt{\omega_s^2+\omega_p^2\pm \sqrt{\omega_s^4-2\omega_s^2\omega_p^2+4\Delta_{ps}\omega_s^2\omega_p^2-\omega_p^4}}\,,
        \label{eq:freq detune}
\end{equation}
where $\omega_p$ and $\omega_s$ represent the WGM and spin resonance frequency, and $\Delta_{ps}$ is the normalised, unitless, inductive mutual coupling \cite{83834} between these resonances. Here, the value of $\Delta_{ps}$ parameterises the rate of coupling $g$ ($g \coloneqq \frac{\Delta_{ps}\omega_{p}}{2}$). This model ignores loss, but suffices to fit the perturbation extremely well in the present study due to the fact that the photon line widths, $\Gamma_{p}=\frac{\omega_{p}}{Q_{p}}$, are of the order kHz, while $g$ is of the order MHz, allowing strong off resonant perturbations of the photon normal modes due to the spins (here $Q_p$ is the WGM $Q$\hyp{}factor). The defect spin concentration ($n$) may be then derived from the rate of coupling, using the following expression;
\begin{equation}
    g=g_L\mu_{B}\sqrt{\frac{\mu_0\omega{_p} n\xi_{\perp}}{4\hbar}}\,,
    \label{eq:spin concentration}
\end{equation}
where $\mu_{B}$ is the Bohr Magneton, $\hbar$ is the reduced Planck's constant, $\xi_{\perp}$ is the perpendicular magnetic filling factor \cite{Jeremy2016} ($\xi_{\perp} \approx 1$ for TM\hyp{}WGMs), $g_L$ was calculated from the data in Fig. \ref{fig:SpecResults}, and $\mu_{0}$ is the permeability of free space.  Note, this assumes the concentration of spins is uniform, and $n$ is the average concentration over the mode volume. The calculation of the defect concentration can thus be extracted by fitting equation \ref{eq:freq detune} to the frequencies of the mode interaction in Fig. \ref{closeup}, and from this fit we estimate the concentration to be $n=8.28\pm1.24\times10^{13}~$cm$^{-3}$.

This study found the presence of dilute spin ensembles in a scheelite sample, with Gd$^{3+}$ being the dominant species. Fe$^{3+}$ and another unknown species ``A" are also shown in the ESR spectra at much lower concentrations. This material exhibits low loss at mK temperatures of order $3\times 10^{-8}$ at $14.934048$ GHz in frequency. The high\hyp{}$Q$s of the WGMs hosted in the cylindrical sample allows one to conduct highly sensitive multi\hyp{}mode spectroscopy with the capability to resolve spin\hyp{}photon interactions with frequency detunings smaller than an order of magnitude below the linewidth of the photon resonances. In principle, this sets the lower bound of observable impurity concentrations on the order of parts per trillion \cite{ppt}. The high coupling rate and high WGM $Q$\hyp{}factor ($\Gamma_p<<2g$) indicate a high cooperitivity. Accordingly, the presence of this dilute spin ensemble may be exploited in multiple avenues from bolometric sensing to qubit applications \cite{PhysRevLett.131.066701,PhysRevB.109.104301} since the low concentration of spins, on the order of ppb, preserve the $Q$\hyp{}factor of the dielectric, whilst exhibiting sensitivity to photon coupling. 

\section*{Acknowledgements}
This work was funded by the ARC Centre of Excellence for Engineered Quantum Systems, CE170100009, and Dark Matter Particle Physics, CE200100008. BM was also funded by the Forrest Research Foundation.
%\newpage

\end{document}